\documentclass[useAMS,usenatbib]{mn2e}
\usepackage{times}
\usepackage{deluxetable}

\usepackage{url}
\usepackage{epsfig}
\usepackage{amsmath}
\usepackage{natbib}
\begin{document}

\title{The Dust \& Gas Properties of M83}

%\author[Foyle et al.]{K.Foyle, C. Wilson, E. Mentuch, G. Bendo, T. Parkin, A. Dariush, M. Baes, M. Boquien, A. Boselli, A. Cooper, J. Davies, S. Eales, S. Madden, M. Smith and L. Spinoglio}
\author[K.~Foyle et al.]
{\parbox{\textwidth}{K.~Foyle,$^{1}$\thanks{E-mail: \texttt{foylek@physics.mcmaster.ca}}
C.~D.~Wilson,$^{1}$ %Mcmaster
E.~Mentuch,$^{1}$ %McMaster
G.~Bendo,$^{2}$ %UK Alma
A.~Dariush,$^{3}$ %Imperial
T.~Parkin,$^{1}$ %McMaster
M.~Pohlen,$^{4}$ %Cardiff
M.~Sauvage,$^{5}$ %CEA
M.~W.~L.~Smith,$^{4}$   %Cardiff
H.~Roussel,$^{6}$ %IAP
M.~Baes,$^{7}$  %GENT
M.~Boquien,$^{8}$ %Amherst
A.~Boselli,$^{8}$ %Marseille
D.~L.~Clements,$^{3}$ %Imperial
A.~Cooray,$^{9}$ %Irvine
J.~I.~Davies,$^{4}$ %Cardiff
S.~A.~Eales,$^{4}$ %Cardiff
S.~Madden,$^{5}$ %CEA
M.~J.~Page,$^{10}$ and
L.~Spinoglio$^{11}$}\vspace{0.4cm}\\   %Italy
\parbox{\textwidth}{$^{1}$Dept. of Physics \& Astronomy, McMaster University, Hamilton, Ontario, L8S 4M1, Canada\\
$^{2}$UK ALMA Regional Centre Node, Jodrell Bank Centre for Astrophysics, School of Physics and Astronomy, University of Manchester, Oxford Road, Manchester M13 9PL, United Kingdom\\
$^{3}$Astrophysics Group, Imperial College, Blackett Laboratory, Prince Consort Road, London SW7 2AZ, UK\\
$^{4}$School of Physics \& Astronomy, Cardiff University, Queen Buildings, The Parade, Cardiff CF24 3A\\
$^{5}$CEA, Laboratoire AIM, Irfu/SAp, Orme des Merisiers, F-91191, Gif-sur-Yvette, France\\
$^{6}$Institut d'Astrophysique de Paris, UMR7095 CNRS, Universit\'e Pierre  \& Marie Curie, 98 bis Boulevard Arago, F-75014 Paris, France\\
$^{7}$Sterrenkundig Observatorium, Universiteit Gent, Krijgslaan 281 S9,  B-9000 Gent, Belgium\\
$^{8}$Laboratoire d'Astrophysique de Marseille, UMR6110 CNRS, 38 rue F.  Joliot-Curie, F-13388 Marseille France\\
$^{9}$Department of Physics \& Astronomy, University of California,
Irvine,CA 92697, USA\\
$^{10}$ Mullard Space Science Laboratory, University College London,
Holmbury St.\ Mary, Dorking, Surrey RH5 6NT\\
$^{11}$Istituto di Fisica dello Spazio Interplanetario, INAF, Via del Fosdso del Cavaliere 100, I-00133 Roma, Italy}}
\maketitle
\label{firstpage}

%\affil{Dept. of Physics \& Astronomy, McMaster University, Hamilton, Ontario, L8S 4M1, Canada}

%\email{foylek@physics.mcmaster.ca}

\begin{abstract}
We examine the dust and gas properties of the nearby, barred galaxy
M83, which is part of the Very Nearby Galaxy Survey.  Using images
from the PACS and SPIRE instruments of {\it Herschel}, we examine the dust
temperature and dust mass surface density distribution.  We find that
the nuclear, bar and spiral arm regions exhibit higher dust
temperatures and masses compared to interarm regions.  However, the distribution of dust
temperature and mass are not spatially coincident.  Assuming a trailing spiral structure, the dust temperature peaks in the spiral arms lie ahead of the dust surface density peaks. The dust mass surface density correlates well with the distribution of molecular gas as traced by CO (J=3$\rightarrow$2) images (JCMT) and the star formation rate as traced by H$\alpha$ with a correction for obscured star formation using 24~$\mu$m
emission.   Using H{\small I} images from THINGS to trace the atomic gas component, we make total gas mass surface density maps and
calculate the gas-to-dust ratio.  We find a mean gas-to-dust ratio of
84 $\pm$ 4 with higher values in the inner region assuming a constant CO-to-H$_{2}$ conversion factor.  We also examine the gas-to-dust ratio using CO-to-H$_{2}$ conversion factor that varies with metallicity.
 \end{abstract}

%\keywords{galaxaes: general ---}
%\doublespace
%\onecolumn
\section{Introduction}
Dust emission represents approximately one third of the bolometric luminosity in normal spiral galaxies and dust acts to shape the spectral energy distribution (SED) of galaxies by attenuating short wavelength emission and re-emitting at longer wavelengths.  Dust can also be used to trace the total gas content in galaxies, since dust and gas often coexist.  An understanding of dust and its distribution are essential to reveal the underlying appearance of galaxies and the processes within (Draine 2003).
  
Until recently it has been difficult to map the distribution and
amount of dust in galaxies because, for most galaxies, only shorter wavelength infrared data has been
available ({\it i.e.} $\lambda$ $<$ 160~$\mu$m).  Without longer wavelengths, only the warmer dust emission is measured and subsequently the dust mass is poorly constrained (Draine et al.\ 2007 and Galametz et al.\ 2011).  While some ground-based telescopes ({\it i.e.} SCUBA on JCMT; Holland et al.\ 1999 and SHARCII on the CSO; Dowel et al.\ 2003) have provided measurements in the far-infrared (350, 450 and 850~$\mu$m) allowing for more accurate dust SED fits ({\it e.g.} Dunne et al.\ 2000, Dunne \& Eales 2000 and Dale et al.\ 2007), these images had low sensitivity due to background noise from the atmosphere.

The {\it Herschel} Space Observatory (Pilbratt et al.\ 2010) has
allowed us to probe the colder dust component in galaxies using
measurements at wavelengths spanning 70 to 500~$\mu$m at high
sensitivity and resolution.  In comparison to the {\it Spitzer} Space
Telescope (Werner et al.\ 2004), which also operated at 70 and
160~$\mu$m (MIPS; Rieke et al.\ 2004), the point spread functions (PSFs) are smaller, the sensitivity is higher and there are fewer latent image effects.

In this work we use {\it Herschel} observations to make dust temperature
and mass maps for M83 which is part of the Very Nearby Galaxy Survey (VNGS;
PI: C.\ Wilson).  Due to its proximity
(4.5 $\pm$ 0.3 Mpc; Thim et al.\ 2003), which allows for high spatial
resolution, and its relatively face-on orientation (inclination angle of 24$^{\circ}$), it makes an excellent candidate to study the spatial distribution of the dust mass and temperature.  

M83 is a starburst galaxy with a prominent bar, which is outlined by dust lanes.  Dynamical studies have shown that gas is funneled along the bar producing high rates of star formation at the centre (Knapen et al.\ 2010).  M83 is rich in molecular gas with 13 per cent of the disk mass consisting of molecular gas (Lundgren et al.\ 2004).  Studies of the spiral arms have shown that tracers of the gas and star formation rate (SFR) are often offset from each other suggesting that the spiral arms dynamically induce star formation (Lord \& Kenney 1991).  However, the situation is complicated in that molecular clouds are also seen coincident with young star forming regions (Rand et al.\ 1999).  Studies of the mid-infrared emission, which traces the very small dust grains and PAHs, have revealed a tight correlation with the radio continuum, which may suggest the anchoring of magnetic fields in the photoionizing shells of clouds (Vogler et al.\ 2005).  The morphology of M83 and the fact that it has coverage at many other wavelengths allows for comparisons of the varying environments within the galaxy and comparisons of the dust distribution to tracers of the gas and SFR.

We estimate the dust temperature and mass of M83 on angular scales of 36 arcsec (790 pc) using 70~$\mu$m and 160~$\mu$m images from the Photodetector Array Camera and Spectrometer (PACS; 
Poglitsch et al.\ 2010) and the 250, 350, 500~$\mu$m maps from the Spectral and Photometric Imaging REceiver 
(SPIRE; Griffin et al.\ 2010).  We compare the spatial distribution of the dust temperature and mass with tracers of the atomic gas, molecular gas and star formation rate for different regions in the galaxy including the nucleus, bar and spiral arms.  In \S2 we describe how the infrared, gas and SFR images were rendered and processed. Our derivation of the dust temperature and dust mass distributions through SED fitting on each pixel is described in \S3 and compared to the distributions of molecular and total gas as well as the star formation rate surface density in \S4. Our conclusions are summarized in \S5.
 
 \section{Observations \& Image Rendering}
 \subsection{Herschel Images}
\begin{figure*}
\centering
 \includegraphics[trim=30mm 5mm 10mm 20mm, clip,angle=270,width=180mm]{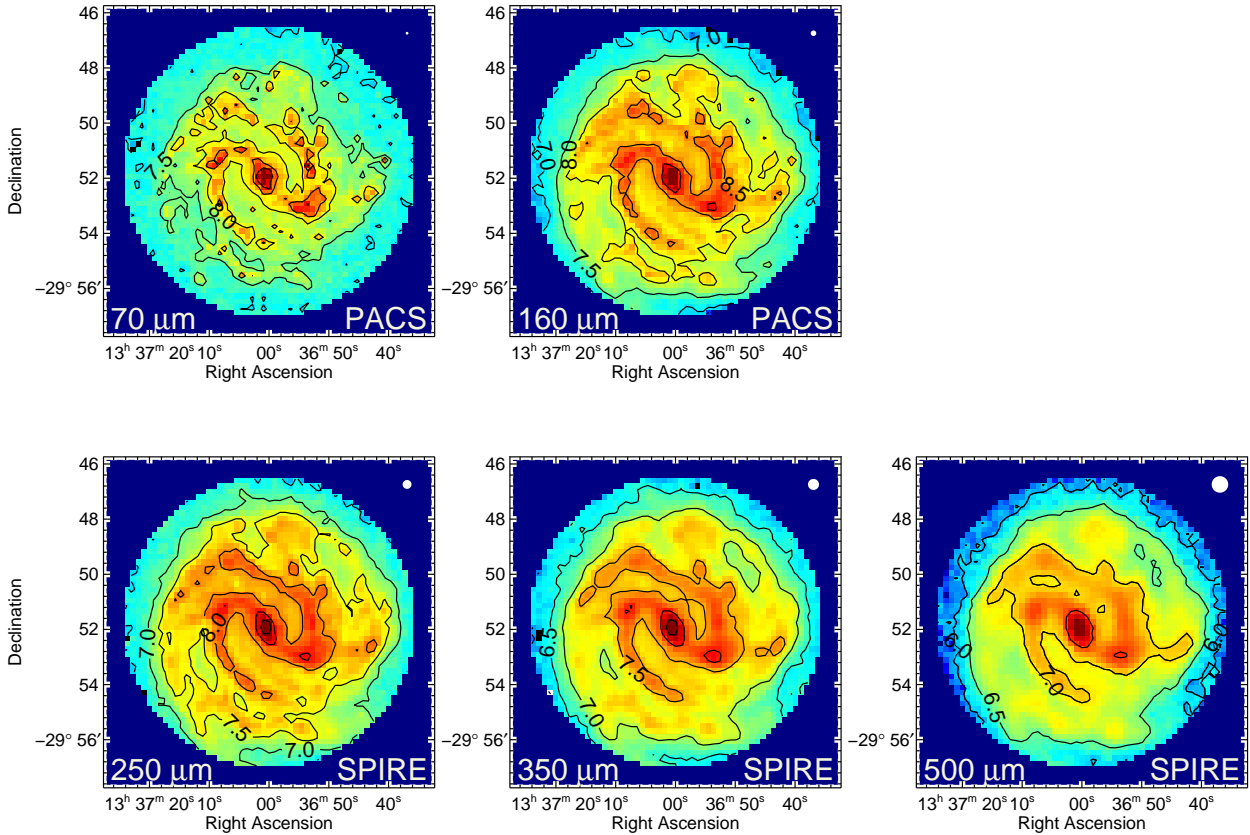}
\caption{M83 at 70~$\mu$m and 160~$\mu$m imaged using PACS (upper panels from left to right) and at 250~$\mu$m, 350~$\mu$m and 500~$\mu$m imaged using SPIRE (lower panel from left to right).  The images are in Jy/sr and are scaled logarithmically with contours at equal increments (6, 6.5, 7, 7.5, 8, 8.5 and 9).  The size of the PSF is illustrated in the top right hand corner of each image (see Table 1 for pixel and PSF sizes).}
\label{all}
\end{figure*}

 We use PACS images that are processed using both \textsc{HIPE} v5 and \textsc{Scanamorphos} v8\footnote[1]{\textsc{Scanamorphos} Documentation: \url{http://www2.iap.fr/users/roussel/herschel/}} (Roussel et al.,
sub.) and SPIRE images processed with \textsc{HIPE} and \textsc{B}ri\textsc{GA}de (Smith et al., in prep).  Details on how the images are processed can be found in Bendo et al.\ (2011), the SPIRE Observer's Manual\footnote[2]{SPIRE Observer's Manual: \url{http://herschel.esac.esa.int/Docs/SPIRE/html/spire_om.html}} and the PACS Observer's Manual\footnote[3]{PACS Observer's Manual: \url{http://herschel.esac.esa.int/Docs/PACS/html/pacs_om.html}}. We briefly describe the steps here.  The PACS images were corrected from the v5 photometric calibration files to v6 (with corrective factors of 1.119 and 1.174 for the 70 and 160~$\mu$m maps respectively) and were converted to Jy/sr from Jy/pixel . The SPIRE images are first converted to Jy/pixel from Jy/beam using the respective beam sizes (423 $\pm$ 3, 751 $\pm$ 4, 1587 $\pm$ 9 arcsec$^{2}$, for the 250, 350 and 500~$\mu$m images respectively).  The images were then converted to Jy/sr.  The SPIRE images were multiplied by the following values to convert from monochromatic intensities of point sources to monochromatic extended sources, 0.9939/1.0113, 0.9898/1.0060 and 0.9773/1.0065 for the 250, 350 and 500~$\mu$m maps respectively.  Table 1 lists the pixel sizes and FWHM of the PSFs for both the PACS and SPIRE images.  Sky subtractions for all five images were made by subtracting the median value of several off-galaxy apertures.  Fig.~\ref{all} shows the five images in their native resolution, before convolution and deprojection.
  %%%%%%%%%%%%%%%%%%%%%%%%%%%%%%%%%%%%%%%%%%%%%%%%%%%%%%%%%%%%%%%%%%%%%%%%%%%%

%%%%%%%%%%%%%%%%%%%%%%%%%%%%%%%%%%%%%%%%%%%%%%%%%%%%%%%%%%%%%%%%%%%%%%%%%%%%
 The images are all aligned and convolved to the 500~$\mu$m images
 using kernels developed by Aniano et al.\ (2011) for PACS and SPIRE.
 The final pixel size is 12 arcsec.  However, in our quantitative analysis we use 36 arcsec pixels, which correspond to the FWHM of the 500~$\mu$m map.  We also deproject the images to a face-on orientation adopting the inclination and position angles from Tilanus \& Allen (1993) (see Table 2).  
 
 The uncertainties for each image are a combination of the uncertainty
 image files, which are corrected for convolution and deprojection,
 sky background uncertainties and the calibration uncertainties (0.03,
 0.05, 0.07, 0.07, 0.07 for the calibration uncertainties of the 70,
 160, 250, 350 and 500~$\mu$m images respectively).

While Multiband Imaging Photometer images from {\it Spitzer} (MIPS;
Rieke et al.\ 2004) are available for M83 at 70~$\mu$m and 160~$\mu$m, they suffer
from latent image effects and saturation in the nucleus due to M83's
brightness (see Bendo et
al.\ 2011 for a discussion).  Thus, we choose not to include these
images in our analysis.

    \begin{table}
\begin{center}
\caption{Characteristics of Images}
\begin{tabular}{ccc}
\hline\hline
Image & Pixel Size [ arcsec] & FWHM of PSF [ arcsec]\\
\hline
70~$\mu$m & 1.4 & 6.0\\
160~$\mu$m & 2.85 & 12.0\\
250~$\mu$m & 6 & 18.2\\
350~$\mu$m & 8 & 24.5\\
500~$\mu$m & 12 & 36.0\\
\hline
\end{tabular}
\end{center}
\end{table}
   \begin{table}
\begin{center}
\caption{Adopted Properties of M83}
\begin{tabular}{cc}
\hline\hline
Inclination & 24$^{\circ}$\\
Position Angle & 225$^{\circ}$\\
Distance & 4.5 Mpc\\
\hline
\end{tabular}
\end{center}
\end{table}

\subsection{H{\small I}, CO \& Gas Maps}
We use 21 cm line emission maps from THINGS (The H{\small I} Nearby Galaxy Survey; Walter et al.\ 2008) to trace the atomic gas. We use the natural weighted maps with a resolution of 13 arcsec and convert
the integrated intensity to a surface density following Leroy et al.\ (2008).  The H{\small I} maps are aligned and convolved to match the 500~$\mu$m images and are deprojected according to the values in Table 2.
 
Muraoka et al.\ (2007) mapped CO (J=1$\rightarrow$0) out to 3.5 kpc.  However, archival CO (J=3$\rightarrow$2) data from the James
Clerk Maxwell Telescope (JCMT; project: M06BN05) with a resolution of 14.5 arcsec allow us to map molecular gas distribution out to 5 kpc. Thus, we use the CO (J=3$\rightarrow$2) but first scale the images to match the CO (J=1$\rightarrow$0) of M83 using the ratio found by Muraoka et al.\ (2007), which varies with radius.  Muraoka et al.\ (2007) found that the ratio of CO (J=3$\rightarrow$2) to CO (J=1$\rightarrow$0)  intensity is as high as 1.0 in the very inner regions and decreases to 0.65 at 0.5 kpc.  Beyond 0.5 kpc the ratio is roughly constant at 0.65.  We use a radially varying ratio, S.  At each radial position, R, between 0 and 0.5 kpc we have
\begin{equation}
S=1.0 - 0.7R\text{  }\text{  }\text{  }\text{  } \text{  }\text{  }\text{  }\text{  }  R < 0.5\text{.}
\end{equation}
Beyond 0.5 kpc we use S=0.65.  Details on how the CO (J=3$\rightarrow$2)  images were processed can be found in Warren et al.\ (2010). We convert from antenna temperature to main beam temperature using a factor $\eta_{MB}=0.6$.   To estimate the surface density of H$_2$ from CO we then use a constant conversion
factor of $X$=2.0$\times$ 10$^{20}$ mol cm$^{-2}$ (K km
s$^{-1}$)$^{-1}$ which corresponds to 3.2 M$_{\odot}$ pc$^{-2}$ (K km
s$^{-1}$)$^{-1}$ (Wilson et al.\ 2009).  We examine a conversion factor that varies with metallicity in \S3.3. The H$_{2}$ maps are aligned
and convolved to match the 500~$\mu$m images and are deprojected according to the values in Table 2.  

Together the H{\small I} and H$_{2}$ maps are combined to make total gas surface density maps using a factor of 1.36 to account for helium.

\subsection{SFR Maps}  
Measuring the SFR requires accounting for both unobscured star formation and dust obscured star formation.  In this work we use H$\alpha$ images, which are corrected for dust obscuration in compact sources using 24~$\mu$m emission maps.
We use continuum subtracted H$\alpha$ images from SINGG (Survey for
Ionization in Neutral Gas Galaxies; Meurer et al.\ 2006) and 24~$\mu$m
observations from Engelbracht et al.\ (2005) that are reprocessed by
Bendo et al.\ (2011, in prep.). The 24~$\mu$m emission traces star
formation in compact regions that are obscured by dust, but also has a
diffuse component from post-main sequence stars.  Here we exclude the
diffuse component and only including the 24$\mu$m emission from
compact sources. Our goal is to trace star forming regions rather
  than radiation associated with star formation.  In Appendix B we
  compare our star formation tracer to linear combinations of images,
  which include the diffuse component.

Several authors have demonstrated that 24~$\mu$m emission from
unresolved sources is correlated with the H$\alpha$ emission from
these sources \citep[e.g.][]{cetal05, cetal07, petal07}.
\citet{cetal07} derived the following equation to use 24~$\mu$m flux
density measurements to correct the H$\alpha$ flux from H{\small II}
regions within a galaxy to produce an extinction-corrected H$\alpha$
flux:
\begin{equation}
f_{H\alpha~corrected} = f_{H\alpha~observed} + (0.031 \pm 0.006)f_{24\mu m}
\label{e_hacorr}
\end{equation}
We use the extinction-corrected H$\alpha$ fluxes given by this
equation as a star formation tracer in our analysis.  However, to
create a map of the extinction-corrected H$\alpha$ emission, we
first identify the unresolved H$\alpha$ sources and their 24~$\mu$m
counterparts.  We then apply the correction to each source and re-map the
extinction-corrected emission into a new map.

To identify the unresolved H$\alpha$ sources, we used SExtractor
\citep[version 2.5.0; ][]{ba96} in double-image mode to create the
initial source catalogues where detection and setting of apertures is
done in the H$\alpha$ image.  All H$\alpha$ images were first
convolved to the resolution of the 24~$\mu$m image using a customised
kernel created using the procedure outlined in
Bendo et al.\ (2011). The images were then regridded to the
pixel size of the 24~$\mu$m map. Sources were selected when at least
20 pixels were $3\sigma$ above the noise level in the H$\alpha$ image
used for detection.  All photometry was measured in apertures with a
40 arcsec ($\sim$27 pixels) diameter.  We used this fixed radius because it
is larger than the peak in the second Airy ring of the 24~$\mu$m data
and the aperture corrections should be less than or equal to the
calibration uncertainties.  To minimize the number of faint spurious
sources in our catalog, we exclude regions beyond the galaxy optical
disk and foreground stars that appear as sources with positive and
negative counterparts in the H$\alpha$ images. We find that when
run SExtractor is run in single-image mode, the mean absolutde value
difference between the 24~$\mu$m
flux densities using the two methods is $\sim$13\%.  Thus, we add a 13\%
systematic error to all of the 24~$\mu$m
measurements.  

After performing these measurements, we corrected the H$\alpha$
measurements for foreground dust extinction within the Milky Way using
$A_R=0.176$, which was calculated by the NASA/IPAC Extragalactic
Database based on data from \citet{sfd98}.  We then applied
Equation~\ref{e_hacorr} to calculate extinction-corrected H$\alpha$
fluxes, remapped the H$\alpha$ fluxes into individual pixels in a map
with very small (1 arcsec) pixels, and then convolved the data with a
Gaussian function with a FWHM of 36 arcsec to match the PSF of the 500~$\mu$m
map for analysis.

Using the calibration of Kennicutt (1998) and Calzetti et al.\ (2007),
we convert the corrected H$\alpha$ maps, which we label as H$\alpha_{,cor}$ into an SFR:
\begin{equation}
\text{SFR}(\text{M}_{\odot}\text{ yr}^{-1}) = 5.3\times10^{-42}\text{ L}_{corr}(\text{H}\alpha_{,cor})\text{ (erg s}^{-1}).
\end{equation}  
For the remainder of this text we will refer to the SFR using this description.  In Appendix B we also compare this tracer to other tracers derived from a linear combination of images.  We examine an SFR made from the combination of the far-ultraviolet (UV) and 24~$\mu$m emission as well as a combination of H$\alpha$ and 24~$\mu$m emission across the whole image without the compact source identification outlined above.

All of the gas, dust and SFR maps are rebinned into 36 arcsec pixels corresponding to a physical size of 790 pc for the purposes of quantitative analysis (see Bendo et al.\ 2011 for a discussion).  36 arcsec represents the FWHM of the convolved data (see Table 1).

\section{Dust Maps}
\subsection{Temperature and Emissivity}
The dust temperature is determined by fitting each pixel in the PACS and SPIRE images with a single modified blackbody function of the form:
\begin{equation}
S_{\lambda}=N B_{\lambda}(T) \lambda^{-\beta},
 \end{equation}
 where $S_{\lambda}$ is the flux density, $B_{\lambda}(T)$ is the Planck function, $N$ is a constant related to the column density of the material and $\beta$ is the emissivity.
  We fit the points using MPFIT, a least-squares curve fitting routine for IDL (Markwardt 2009).  Because the dust emission does not have a flat spectrum across the bands, during our fitting routine, we apply colour corrections to the five images. The appropriate colour corrections were determined by first fitting the SED of the five images and determining the spectral index or slope at each point.  The photometric colour corrections were determined by interpolation from the tables provided in the observer manuals of PACS and SPIRE.  This process was repeated iteratively.  

The grain emissivity parameter ($\beta$) of the dust reflects its properties and is
important for determining the dust temperature and mass.  Typically,
it has a value of $\sim$2 (Draine \& Lee 1984).  However, it can vary
from as low as 1 to as high as 3 (Shetty et al.\ 2009a and references
therein).  While constant values of $\beta$ of either 1.5 or 2 have
been used in many studies, $\beta$ and the dust
temperature may be anti-correlated due to the properties of amorphous
grains (Dupac et al.\ 2001; Meny et al.\ 2007; Paradis et al.\ 2010).
Thus, fixing $\beta$ can lead to erroneous temperature
values. However, noise and line-of-site temperature variations can
also produce this inverse relation making uncovering the intrinsic
relation between $\beta$ and temperature challenging (Shetty et
al.\ 2009a; Shetty et al.\ 2009b; Malinen et al.\ 2010).

\begin{figure}
\centering
 \includegraphics[angle=-90,width=140mm]{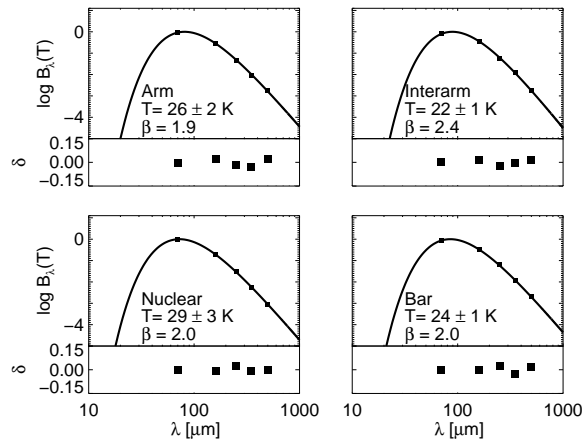}

\caption{Sample modified blackbody fits for regions in the nucleus, bar, arm and interarm with the colour-corrected PACS and SPIRE points using a variable $\beta$ (see centre of each panel for fit value).  The peak of the function determines the temperature (listed in each panel).  The functions are shown in log-log and are normalized to the peak of the function.  The uncertainties are smaller than the symbol sizes.  The deviation between each point and the fit ({\it i.e.}\ $\log S_{\lambda} - \log I_{\lambda}$) is shown below each panel.  }
\label{sed}
\end{figure}

In this work we fit the dust emission using both a constant and a
variable value of $\beta$.  For a variable $\beta$ we permit values
between 1.0 (the limit of the Kramers-Kronig dispersion relation) and
2.5.  The mean value of $\beta$ over the entire galaxy is 2.1$\pm$0.1.
For our constant $\beta$ fit, we assume a value of 2.0, which is close
to the mean value found when $\beta$ was allowed to vary.  The mean
$\chi^{2}$ over all the fit pixels when using a variable $\beta$ is
0.5 with a standard deviation of 0.3.  In the case of a constant
$\beta$ the mean value is 0.7 with a standard deviation of 0.3. These $\chi^{2}$ values are small suggesting that our errors may be large.  This is
likely due to the calibration errors, which are correlated.  Fig.~\ref{sed} shows four sample fits in log-log space for pixels in the nuclear, bar, spiral arm and interarm regions.  The curves are all normalized to the peak value.  The bottom panel shows how the data deviates from the fit ({\it i.e.}\ $\log S_{\lambda} - \log I_{\lambda}$).  In each panel we list the fitted temperature and emissivity. 

\begin{figure*}
\centering
 \includegraphics[trim=30mm 10mm 10mm 10mm, clip,angle=-90,width=180mm]{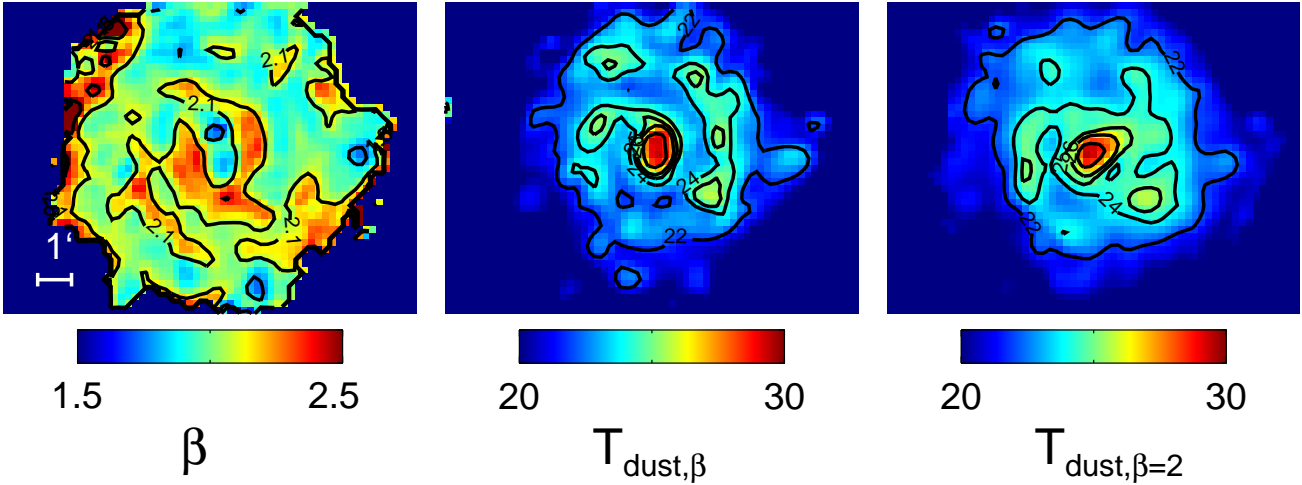}

\caption{Spatial distribution of $\beta$ (left), dust temperature using a variable $\beta$ in the modified blackbody fit (centre) and dust temperature using a constant $\beta$ of 2 in the modified blackbody fit (right).  All three images show the same field of view.  For scale we show a 1$'$ distance on the deprojected images (left panel).  The contours in the $\beta$ map are at 1.5, 1.8, 2.1 and 2.4.  The contours in the two temperature maps (centre and right panels) are the same and lie between 22 and 30 K in increments of 2.  Over much of the galaxy the variable $\beta$ is roughly 2.  However, there is an inner ring of higher $\beta$ values which leads to the differences in the central regions of the two temperature maps.}
\label{tempbeta}
\end{figure*}

Fig.~\ref{tempbeta} shows the spatial distribution of $\beta$ when it
was allowed to vary (left) and the resulting temperature maps for a variable
$\beta$ (centre) and constant, $\beta=2$ (right).   The high $\beta$ values on the edge of the map are likely due to the low signal-to-noise in these regions.  One notes clear differences in the two temperature maps, particularly in the inner regions.  With a constant $\beta$, the central temperature peak is elongated perpendicular to the bar and shows an outer ring-like structure, whereas with a variable $\beta$, the central peak is much more symmetrical.  Over much of the galaxy, the value of $\beta$ was close to 2, but in the inner regions there is a ring of higher $\beta$ values.  This ring is the source of the differences in the inner regions of the two temperature maps.  Bendo et al.\ (2003) also found values of $\beta$ close to 2 in the inner regions (within 135 arcsec) of M83 using ISO (Kessler et al.\ 1996) and SCUBA (Holland et al.\ 1999) data.  M83 had the steepest emissivity law in their analysis of eight galaxies using a variable $\beta$.  

\begin{figure}
\centering
 \includegraphics[trim=10mm 10mm 10mm 10mm, clip,angle=-90,width=90mm]{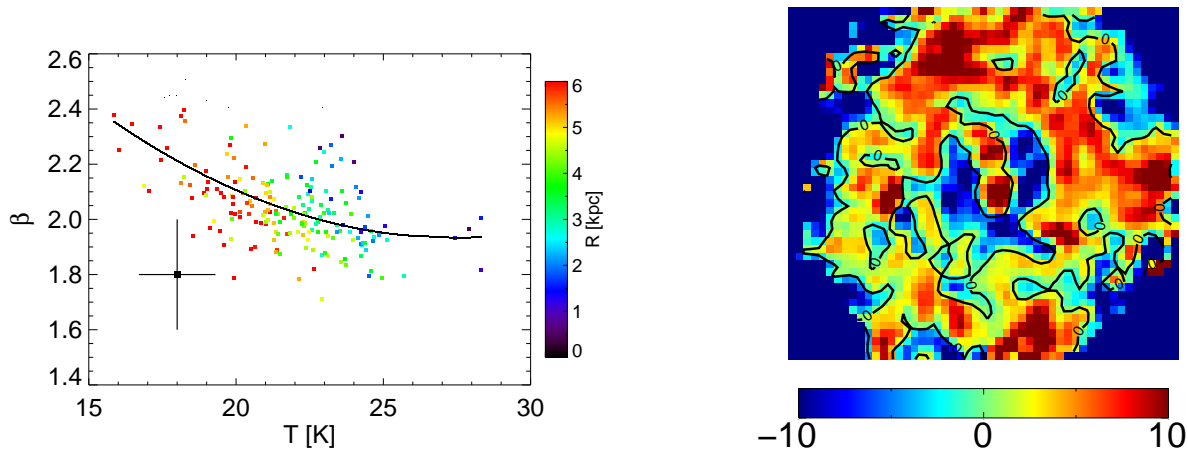}

\caption{$\beta$ versus temperature for the variable $\beta$ SED
  fits. Each point represents a 36 arcsec region corresponding to the
  FWHM of the PSF in the convolved maps. The colours mark the radial
  position of the points with blue and green points within 3 kpc of
  the centre and yellow and red points extending further out.  The
  black point illustrates the mean uncertainties of $\beta$ and the
  temperature.  The solid line shows the two-degree polynomial curve
fit to the points, taking into consideration the uncertainties.  We find that $\beta$ and the temperature are anti-correlated in agreement with other studies.  It is unclear whether this anti-correlation is intrinsic or is simply a reflection of noise and line-of-sight temperature variations (Shetty et al.\ 2009a and b). }
\label{tbscatter}
\end{figure}

Fig.~\ref{tbscatter} shows how $\beta$ varies with temperature. There is an anti-correlation between $\beta$ and
temperature.  This is in agreement with other studies ({\it i.e.}\ Dupac et al.\ 2001,
Paradis et al.\ 2010 and Dariush et al.\ 2011) and recent results from {\it Planck}, which have revealed an anti-correlation between $\beta$ and temperature from SED fits of the integrated emission of nearby galaxies (Planck Collaboration et al.\ 2011).  

However, whether we have uncovered the
intrinsic inverse relation remains unclear as noise and line-of-sight
temperature variations can also produce a similar relation (Shetty et
al.\ 2009a and 2009b).  While there is a trend of decreasing $\beta$
with increasing temperature, we note that in the inner regions (dark blue
points) where there are mostly high temperatures (T $>$ 25 K), there are some regions with $\beta$ exceeding 2.  These values lie in the ring-like structure seen in the left panel of Fig.~\ref{tempbeta} and deviate from the anti-correlation.  While it is possible that the ring is simply an artifact of imperfect PSF matching, it may also represent a region of different physical properties.  

Using a single temperature modified blackbody fit means that at each pixel we have an average dust temperature. Typically, galaxies have a mixture of dust at different temperatures with a warmer and cooler component.  Here we briefly examine to what extent these features might affect our results.  In order to examine the possibility of a warmer component, we fit the SED with our constant $\beta$ value of 2 dropping the 70~$\mu$m point.  If the 70~$\mu$m point is higher than our fit using 160, 250, 350 and 500$\mu$m, this suggests a warmer component ({\it i.e.}\ Bendo et al.\ 2010a and Smith et al.\ 2010).  We find that the SED fit over-predicts the 70~$\mu$m value over much of the galaxy.  This suggests that including the 70~$\mu$m point is necessary in order to properly constrain the peak.   By examining the far-infrared colour variations in M83, Bendo et al.\ (2011)  found that the dust could be divided into a component radiating primarily at 250-500~$\mu$m and another at 70-160~$\mu$m, which is not evident in the test we perform here.  

\begin{figure}
\centering

 \includegraphics[trim=20mm 50mm 0mm 20mm, clip,angle=-90,width=100mm]{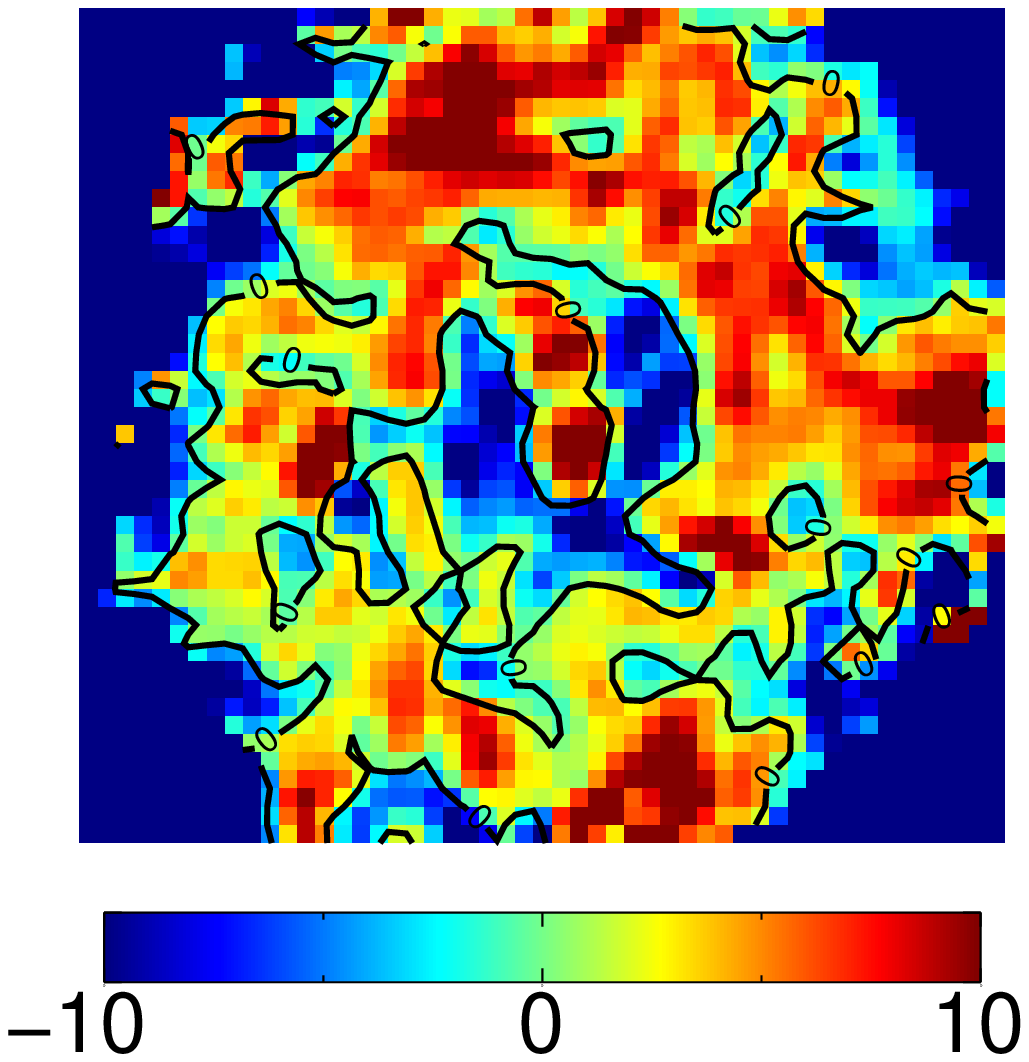}
\caption{The percent difference between the flux of the 500~$\mu$m map and that predicted by the fit using the 70, 160, 250 and 350~$\mu$m maps.  Positive values mean the observed value is greater than predicted, while negative values mean the observed is less than predicted.  With the exception of the inner ring, much of the galaxy shows an excess of 500~$\mu$m emission.}
\label{dif}
\end{figure}

There may also be a cold dust component (T$\sim$5-10K), which has been attributed to a so-called "submm excess" ({\it i.e.}\ Galliano et al.\ 2005, Galliano et al.\ 2005, Bendo et al.\ 2006, Zhu et al.\ 2009, O'Halloran et al.\ 2010 and Gordon et al.\ 2010).  This submm excess has also been explained, in the case of the Milky Way, by wavelength-dependent variations in the dust emissivity ({\it i.e.}\ Reach et al. 1995 and Paradis et al.\ 2009).  Typically, one requires measurements at longer wavelengths ({\it i.e.}\ 850~$mu$m) in order to identify the excess, but even at 500~$\mu$m the excess may present itself ({\it i.e.}\ Gordon et al.\ 2010).  We repeat a similar procedure to that outlined above for the 70~$\mu$m observations.  In general, we find that the 500~$\mu$m observed value exceeds the predicted value of the SED fit using the other four wavelengths by $\sim$10 per cent.  This is in agreement with the excess found by Gordon et al.\ (2010).   Fig.~\ref{dif} shows the percent difference between 500~$\mu$m (right) point and the fit using the other four maps.  Positive values indicate an excess.  We note that the regions where the 500~$\mu$m is under-predicted correspond to the ring-like structure in the variable $\beta$ map.  Since we were forced to keep $\beta$ fix in order to properly fit the SED with only four points, this ring may be an artifact of fixing $\beta$.  We stress that the excess found here is small in comparison to our uncertainties and longer wavelength measurements are necessary to confirm this finding.

   %%%%%%%%%%%%%%%%%%%%%%%%%%%%%%%%%%%%%%%%%%%%%%%%%%%%%%%%%%%%%%%%%%%%%%%%%%%%

%%%%%%%%%%%%%%%%%%%%%%%%%%%%%%%%%%%%%%%%%%%%%%%%%%%%%%%%%%%%%%%%%%%%%%%%%%%%

    %%%%%%%%%%%%%%%%%%%%%%%%%%%%%%%%%%%%%%%%%%%%%%%%%%%%%%%%%%%%%%%%%%%%%%%%%%%%

%%%%%%%%%%%%%%%%%%%%%%%%%%%%%%%%%%%%%%%%%%%%%%%%%%%%%%%%%%%%%%%%%%%%%%%%%%%%
    %%%%%%%%%%%%%%%%%%%%%%%%%%%%%%%%%%%%%%%%%%%%%%%%%%%%%%%%%%%%%%%%%%%%%%%%%%%%

%%%%%%%%%%%%%%%%%%%%%%%%%%%%%%%%%%%%%%%%%%%%%%%%%%%%%%%%%%%%%%%%%%%%%%%%%%%%

    %%%%%%%%%%%%%%%%%%%%%%%%%%%%%%%%%%%%%%%%%%%%%%%%%%%%%%%%%%%%%%%%%%%%%%%%%%%%

%%%%%%%%%%%%%%%%%%%%%%%%%%%%%%%%%%%%%%%%%%%%%%%%%%%%%%%%%%%%%%%%%%%%%%%%%%%%
\subsection{Dust Mass}
We calculate the dust mass using the following equation at  250~$\mu$m:
\begin{equation}
M_{dust}=\frac{S_{250}D^{2}}{\kappa_{250} B(\lambda_{250},T)}\text{ ,}
\end{equation}
 where $S_{\lambda}$ is the flux density from our SED fit at our chosen 250~$\mu$m
 wavelength ({\it i.e.}\ not the value in the 250~$\mu$m map), D is the distance to M83 (4.5 Mpc; Thim et al.\ 2003),
 $B(\lambda,T)$ is the Planck function using 250~$\mu$m and the best fit
 temperature and $\kappa$ is the assumed emissivity.  We use a
   $\kappa$ of 2.72$\times$10$^{5}$ ($\lambda/ \mu$m)$^{-2}$
 cm$^{2}$g$^{-1}$ where $\lambda$=250~$\mu$m.  The value of $\kappa$ is taken from Li \& Draine (2001) and
 modified with a correction discussed in Draine (2003).  

We calculate the mass at 250~$\mu$m as it is not overly sensitive to the
 temperature like the shorter PACS wavelengths and has lower
 uncertainties than the other SPIRE wavelengths.  We tested the other wavelengths as fiducial wavelengths for the dust mass maps and found that the total dust masses agree within the uncertainties.  
  
We compared the spatial distribution of the dust surface density using either a constant $\beta$ or
 variable $\beta$ and found that they are very similar. We find a total dust mass of 4.0 $\pm$ 0.6 $\times$ 10$^{7}$ M$_{\odot}$ with a variable $\beta$ and 3.9 $\pm$ 0.6 $\times$ 10$^{7}$ M$_{\odot}$ with a constant $\beta$ within r=300 arcsec. The difference between the two dust mass maps in individual pixels is at the 10 per cent level.  Galametz et al.\ (2011) did global SED fits of M83 using infrared data from IRAS and submm data at 540~$\mu$m from Hildebrand et al.\ 1977.  Without the submm data, they found a total dust mass of 8.5 $\times$ 10$^{8}$ M$_{\odot}$.  However, when they included the submm data, the total dust mass decreased to 8.5 $\times$ 10$^{6}$ M$_{\odot}$, illustrating the need for longer wavelength data in order to properly constrain the SED fit.  Our dust mass, while higher, is much closer to their global fit using the older longer wavelength data.  We also compared the total dust mass from the fit, without using the 70~$\mu$m and 500~$\mu$m maps.  In these cases, we assumed a $\beta$ value of 2.  In both cases, the total dust mass was identical to that using all five maps within the uncertainties.  In Appendix A we also compare our dust mass estimations with those found using the models of Draine \& Li (2007).

 \section{Results \& Discussion}
 \subsection{Distribution of Dust Temperature and Mass} 
\begin{figure*}
\centering
 \includegraphics[trim=10mm 10mm 10mm 10mm, clip,angle=-90, width=150mm]{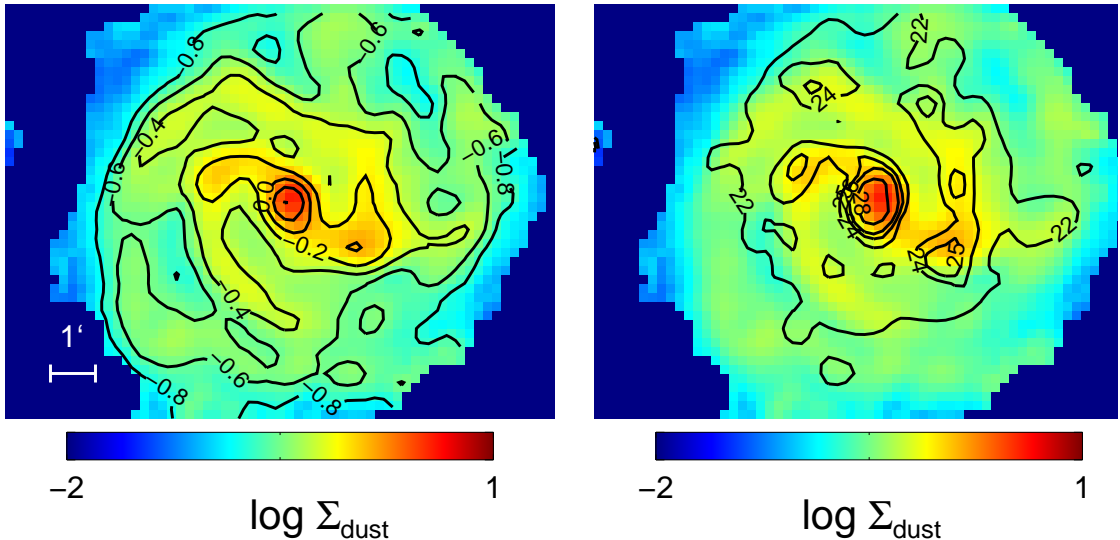}
 
\caption{Dust mass surface density map in log M$_{\odot}$ pc$^{-2}$ for M83 with contours (left) and with contours of the temperature map overlaid (right).  The temperature map is that made with a variable $\beta$ (see middle panel of Fig.~\ref{tempbeta}).  In many cases, peaks in temperature are offset ahead of peaks in mass in the spiral and bar structures.}
\label{temp}
\end{figure*}

 Figs.~\ref{tempbeta} and Fig.~\ref{temp} show the dust temperature and dust mass surface density maps respectively for M83 based on the modified blackbody fits with a variable $\beta$.  The highest temperatures ($\sim$ 30K) are found in the central peak, the spiral arms and the southwestern bar-spiral arm transition.  This is likely due to the increased star formation activity in these regions and, in the case of the bar end, orbit crowding of the molecular gas (Kenney \& Lord 1991) and subsequent star formation activity. Regions outside of the nucleus, but along the bar show lower temperatures, which are comparable to the temperatures in the interarm regions.  
 
The dust mass surface density is shown in Fig.~\ref{temp} in log solar masses per square parsec.  The regions of highest dust concentration are located in the nucleus and at the ends of the bar.  The right panel shows the dust surface density with the temperature contours overlaid. The peak temperature in the south-western bar-spiral arm transition peak and the north-eastern spiral arm are offset from the dust mass surface density peaks.  The dust temperature is highest ahead of the dust mass surface density arms if we assume a trailing spiral structure rotating in the clockwise direction.  Dust temperature and mass maps for  NGC 4501 and NGC 4567/8  in the Herschel Virgo Cluster Survey have also shown that the peaks in dust temperature were not coincident with those in mass (Smith et al.\ 2010).
   %%%%%%%%%%%%%%%%%%%%%%%%%%%%%%%%%%%%%%%%%%%%%%%%%%%%%%%%%%%%%%%%%%%%%%%%%%%%

%%%%%%%%%%%%%%%%%%%%%%%%%%%%%%%%%%%%%%%%%%%%%%%%%%%%%%%%%%%%%%%%%%%%%%%%%%%%
\subsection{Comparison with Gas \& SFR}
\begin{figure*}
\centering
 \includegraphics[trim=10mm 10mm 10mm 10mm, clip,angle=-90,width=170mm]{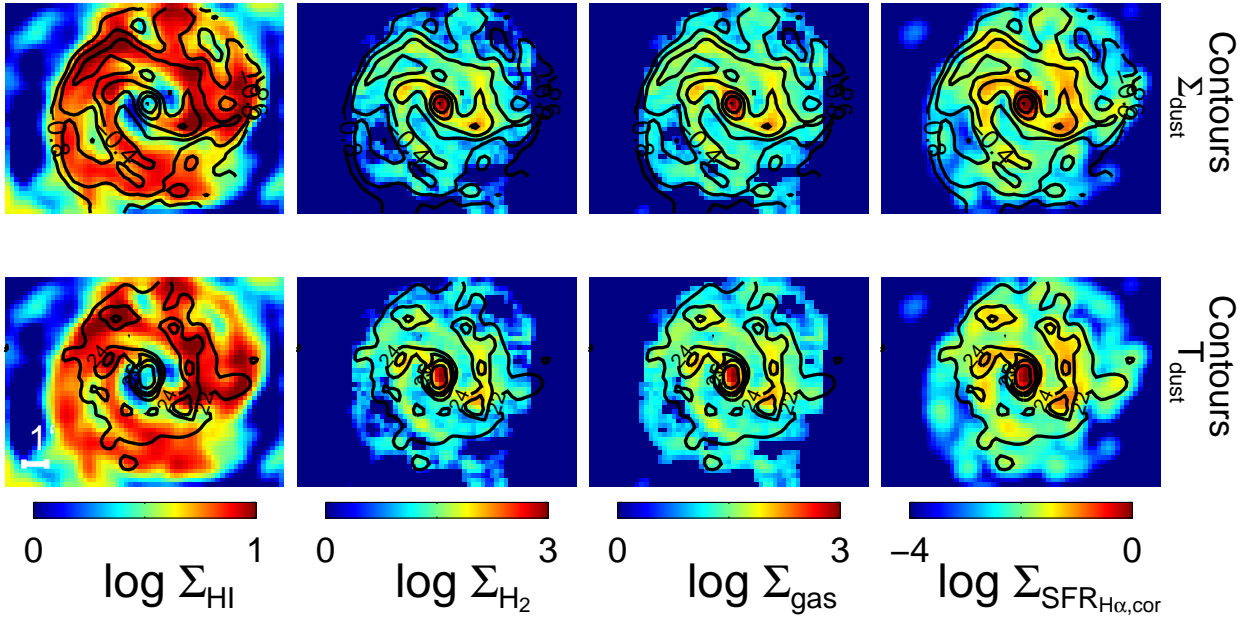}
 
\caption{Atomic (left), molecular (second from left) and total (third from left) gas surface density (logarithmic scale) in M$_{\odot}$ pc$^{-2}$ and the SFR (right) as determined from dust-corrected H$\alpha$ emission (as described in \S2.3) in M$_{\odot}$ kpc$^{-2}$ yr$^{-1}$ .  Contours of the dust mass surface density (upper panel) and temperature (bottom panel) distribution are overlaid.  We find close agreement between the dust mass surface density and molecular gas and SFR distribution.}
\label{gas}
\end{figure*}

\begin{figure*}
\centering
 \includegraphics[trim=10mm 0mm 10mm 10mm, clip,angle=-90,width=170mm]{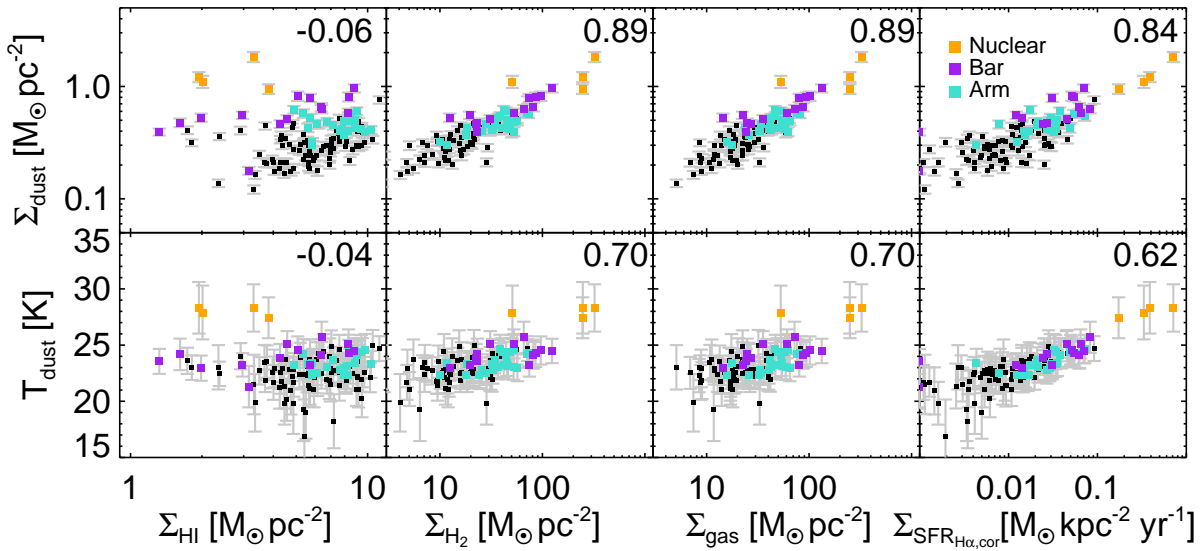}
 
\caption{The dust temperature (lower panel) and dust mass surface density (upper panel) versus atomic gas (left), molecular gas (second from left), total gas (third from left) mass and the SFR (right) as traced by dust-corrected H$\alpha$ emission.  Each point represents a pixel 36 arcsec in size (the FWHM of the convolved maps).  The pixels in the nucleus, bar and spiral arms are denoted in orange, purple and cyan respectively.   The black pixels are in the interarm region or are beyond the extent of the clearly defined spiral structure.  The correlation coefficients between the different tracers are listed in the upper right of each panel.  The dust mass surface density is tightly correlated with the molecular gas, total gas and SFR.  The dust temperature is surprisingly only somewhat correlated with the SFR.  While a rise in temperature is associated with increased molecular gas and SFR, the atomic gas mass is associated with a decrease in temperature. }
\label{gas_dust}
\end{figure*}

We now compare the spatial distribution of the dust mass surface density and temperature with tracers of the atomic and molecular gas and the SFR.  In Fig.~\ref{gas} we show the tracers with contours of the dust surface density (upper panel) and contours of the dust temperature (lower panel).  In Fig.~\ref{gas_dust} we plot the pixels in the dust mass surface density (upper row) and temperature maps (lower row) versus those in the different tracers.  Each point represents an area of 36 arcsec (the FWHM of convolved beam) and we denote regions in the nucleus, bar and
spiral arms in orange, purple and cyan respectively.  The choice for
these regions was made by-eye using 3.6~$\mu$m images from IRAC
(Spitzer Local Volume Legacy; Dale et al.\ 2009) that were convolved to the 500~$\mu$m resolution and deprojected.  3.6~$\mu$m maps roughly trace the underlying stellar mass density, which makes them well-suited for probing morphological structures.  Although PAH and hot dust emission can also contribute in this band in regions dominated by emission associated with star formation (Mentuch et al.\ 2010), since these regions in M83 already correspond to regions with enhanced stellar mass densities, this contamination does not affect our qualitative approach in decomposing the galaxy's morphology.

Examining the upper panel of Fig.~\ref{gas} we see that peaks in the dust mass surface density are virtually coincident with those of the molecular gas and the SFR.  The position of the spiral arm structure in the dust mass surface density is also coincident with that in the atomic gas map.  In the nucleus, however, we find little atomic gas.  In contrast to the excellent spatial correspondence between the gas and SFR tracers with the dust mass surface density, we note that in the lower panel of Fig.~\ref{gas} the dust temperature does not have as close of an agreement.

In Fig.~\ref{gas_dust} we examine in a more quantitative fashion the
correlations between the dust and the different gas and SFR tracers
(see the correlation coefficients in the upper right of each panel).
Again here we find that the molecular gas is tightly correlated with
the dust mass surface density.  While the total gas is also tightly
correlated with the dust surface density, it is important to note,
that over the radii considered, the molecular component dominates the
mass budget (Crosthwaite et al.\ 2002) so this is not too surprising.
We note that the dust temperature tends to increase with rising
molecular gas and SFR, but remains flat with the atomic gas density.  This is simply the result of the increasing atomic gas density with radius, but decreasing dust temperature with radius (see Fig.~\ref{radgas} for radial profiles).  The dust temperature is somewhat correlated with the molecular gas and SFR, but the correlations are not as strong as with the dust mass surface density.   Bendo et al.\ (2011) has also found offsets between colour temperatures (160/250 and 250/350) with the H$\alpha$ spiral arm structure.  This may reveal that while the dust mass surface density is tightly correlated to molecular clouds and thus, star formation, dust heating may be caused by a variety of sources and not just recent star formation activity. Stars form in dense, cold molecular clouds.  After the stars have formed, they heat the surrounding dust that lies within one mean free path.  Thus, it is possible that colder dust lies in the star forming regions.  As the dust disperses it will be heated by more stars in the surrounding region and will subsequently become hotter.

Examining the different regions of the galaxy reflected by the colour-coding of the points, we find that the nuclear regions tend to have higher dust temperatures and higher dust masses for the same value of the gas surface density.  The latter, however, may be due to radial variations in the gas-to-dust ratio related to metallicity (see \S4.5 for a further discussion of this).

  %%%%%%%%%%%%%%%%%%%%%%%%%%%%%%%%%%%%%%%%%%%%%%%%%%%%%%%%%%%%%%%%%%%%%%%%%%%%

%%%%%%%%%%%%%%%%%%%%%%%%%%%%%%%%%%%%%%%%%%%%%%%%%%%%%%%%%%%%%%%%%%%%%%%%%%%%
   
  %%%%%%%%%%%%%%%%%%%%%%%%%%%%%%%%%%%%%%%%%%%%%%%%%%%%%%%%%%%%%%%%%%%%%%%%%%%%

%%%%%%%%%%%%%%%%%%%%%%%%%%%%%%%%%%%%%%%%%%%%%%%%%%%%%%%%%%%%%%%%%%%%%%%%%%%%

\subsection{Radial Profiles}
\begin{figure*}
\centering
 \includegraphics[trim=0mm 0mm 0mm 130mm, clip,angle=-90,width=80mm]{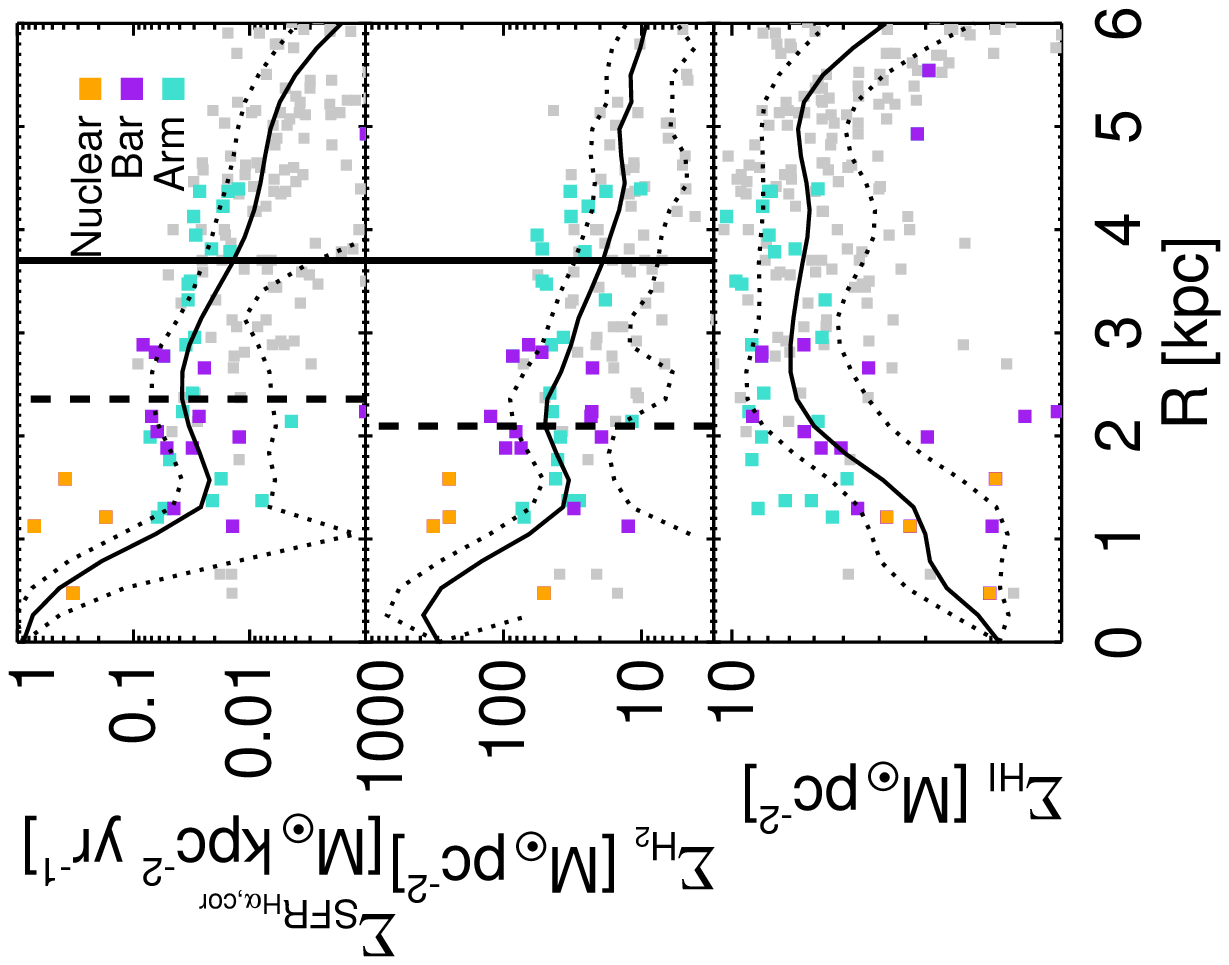}
\includegraphics[trim=0mm 0mm 0mm 130mm, clip,angle=-90,width=80mm]{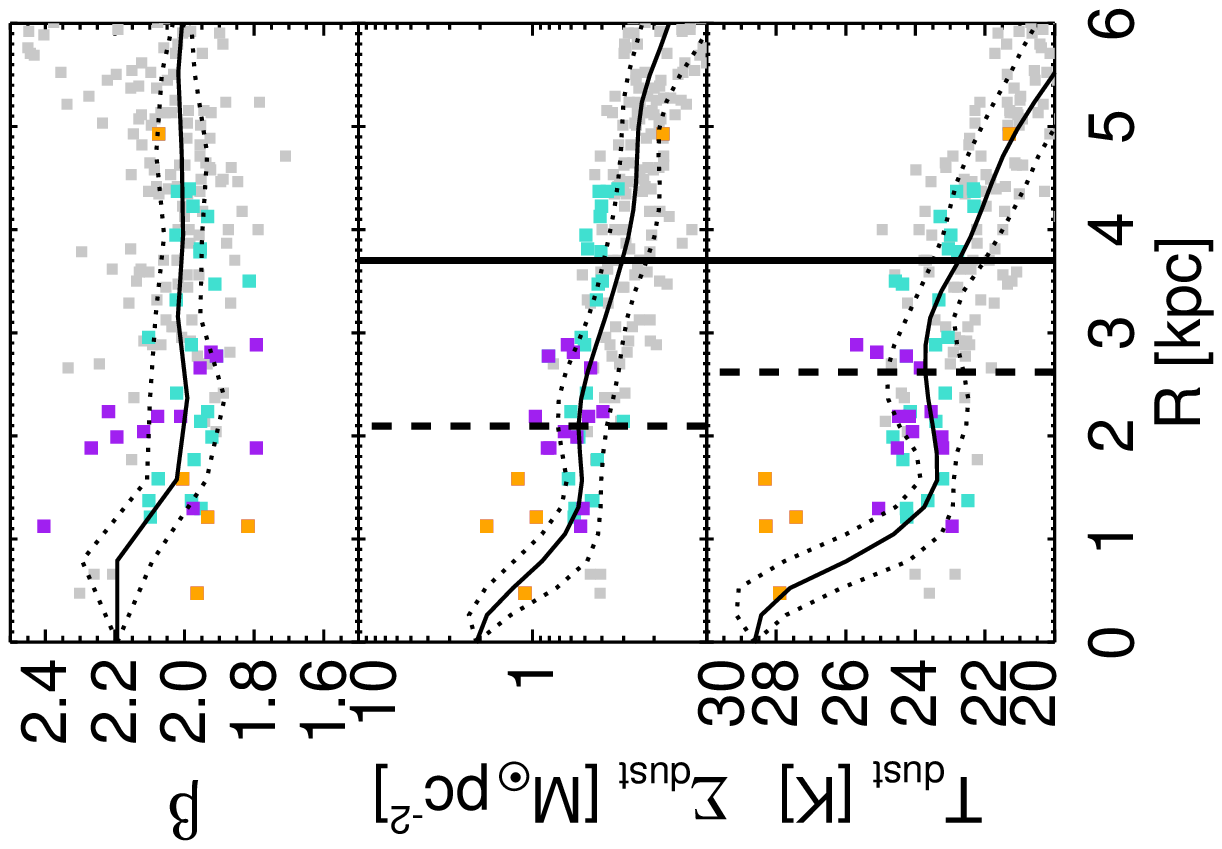}
\caption{Radial profiles of the atomic gas (left, bottom), molecular
  gas (left, middle), SFR (left, top), dust temperature (right,
  bottom), dust mass surface density (right, middle) and emissivity,
  $\beta$ values (right, top).  The solid curve denotes the azimuthal
  average and the dotted lines show the standard deviation of this
  average.  36 arcsec pixels in the nuclear, bar and spiral regions
  are marked in orange, purple and cyan respectively.  Corotation and
  the shoulder position (vertical solid and dashed line respectively) are marked in the molecular gas, SFR, dust temperature and mass surface density panels. }
\label{radgas}
\end{figure*}

Fig.~\ref{radgas} shows the radial profiles of the gas tracers, SFR, dust temperature, dust mass surface density and dust emissivity profiles.  Here again we mark the pixels in the nuclear (orange), bar (purple) and arm (cyan) regions based on the 3.6~$\mu$m map.  The solid line denotes the azimuthal average using radial annuli of 12 arcsec width.  The dashed lines show the standard deviation around the azimuthal average.  The points represent 36 arcsec pixels.

The molecular gas profile of M83 (centre panel of Fig.~\ref{radgas}) is typical of a strongly barred spiral galaxy.  The profile shows a central peak of molecular gas that declines, but then rises again producing a ``shoulder".  The shoulder forms at the bar-spiral arm transition region (Nishiyama et al.\ 2001) and is also seen in the SFR, dust temperature and dust mass surface density profiles.  In investigating the UV extinction profiles of M83, Boissier et al.\ (2005) also found that the SFR and extinction rates were enhanced at the bar-spiral arm transition region. 

We identified the peak of the shoulder in the different profiles by finding the first maximum beyond 1 kpc.  It is marked in each plot by the dashed vertical line and it can also be seen in the transition from purple to cyan squares.  The shoulder does not seem to be associated with the position of corotation, which is beyond the bar end for M83 and is marked in these plots with the solid vertical line at 3.7 kpc (Lundgren et al.\ 2004).  

The shoulder is associated with high dust temperature and mass surface densities in the bar-spiral arm transition (especially the southwestern arm) as seen in the spatial distribution maps in \S4.1.  The position of the shoulder varies with the different profiles (see Table 3 for the positions).  The shoulder in the molecular gas and dust mass surface density profile are coincident.  As we discussed in \S4.3, there is close spatial agreement between the two maps.  Both the dust temperature and SFR have shoulders that are further out in the disk, and surprisingly they are not coincident.  

  \begin{table}
\begin{center}
\caption{Shoulder Position}
\begin{tabular}{cc}
\hline\hline
Profile & Shoulder Position [kpc] \\
\hline
Dust Temperature & 2.9\\
Dust Mass & 2.1\\
SFR & 2.4 \\
Molecular Gas & 2.1\\
\hline
\end{tabular}
\end{center}
\end{table}

We note in the bottom right panel Fig.~\ref{radgas} that the temperature of the dust in the bar region right before the shoulder is much higher than average (denoted by solid curve).  This is likely due to increased SF due to orbit crowding in this region.

The top right panel of Fig.~\ref{radgas} shows how the value of
$\beta$ varies with radius.  $\beta$ is close to 2 over much of the
disk, with higher values in the inner 1 kpc.

     %%%%%%%%%%%%%%%%%%%%%%%%%%%%%%%%%%%%%%%%%%%%%%%%%%%%%%%%%%%%%%%%%%%%%%%%%%%%

%%%%%%%%%%%%%%%%%%%%%%%%%%%%%%%%%%%%%%%%%%%%%%%%%%%%%%%%%%%%%%%%%%%%%%%%%%%%

\subsection{Gas-to-Dust Ratio}
\begin{figure}
\centering
 \includegraphics[trim=10mm 20mm 5mm 10mm, clip,angle=-90,width=100mm]{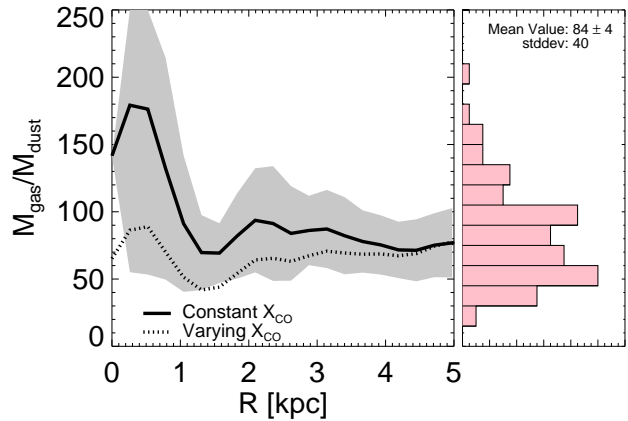}

\caption{GDR versus radius (left) and histogram of the GDR (right).  The azimuthal average of the GDR using a constant X$_\mathrm{CO}$ factor (solid line) and a radially varying one (dashed line) are shown.  The grey shaded region shows the area within one standard deviation of the azimuthal average. The histogram is normalized to the peak value and the mean GDR and standard deviation are listed (upper right).}
\label{gdr}
\end{figure}

Fig.~\ref{gdr} shows the radial variation of the gas-to-dust ratio
(GDR), which is derived using the total gas mass (atomic and
molecular) divided by the dust mass.  Both the azimuthal average
(black solid line) and the region within the standard deviation (grey
shaded area) are shown.  Across M83 we find a mean value of 84 $\pm$ 4 (standard deviation: 40) which is roughly the galactic value of 100-200 ({\it e.g.}\ Sodroski et al.\ 1997).  The right vertical panel of Fig.~\ref{gdr} shows the histogram of the GDR.  Only a few regions in the galaxy have high GDRs in contradiction with previous estimates for M83 (Devereux \& Young 1990) which did not probe the cold dust.  

In the central regions, the GDR attains much higher values (over 200).  This may simply be due to higher metallicities in the inner regions.  One would expect a higher GDR with increasing metallicity.   It may also  be due to the X$_\mathrm{CO}$ factor which can often overestimate the molecular gas in the central regions of spiral galaxies including the Milky Way by a factor of 2 or more ({\it i.e.} Regan 2000).  

The X$_\mathrm{CO}$ factor has been shown to be anti-correlated with metallicity in several studies ({\it i.e.}\ Wilson 1995, Arimoto et al.\ 1996, Boselli et al.\ 2002 and Israel 2005).  A recent study by Magrini et al.\ (2011) has shown that using an X$_\mathrm{CO}$ factor that varies with metallicity may have a large impact on the radial profile of the GDR.  This is particularly true for the inner regions, where the molecular gas dominates. 

In light of this, we re-calculate our estimate of molecular hydrogen using an X$_\mathrm{CO}$ factor which varies with metallicity.  Israel (2000) has derived a relationship between the X$_\mathrm{CO}$ factor and the oxygen abundance as:

\begin{equation}
\mathrm{log} X_\mathrm{CO}= 12.2 -2.5 \mathrm{ log} \frac{O}{H}
\end{equation}

We use the oxygen abundance gradients of Bresolin et al.\ (2009), which are based on linear regression fits of the results of several strong-line studies in the outer regions of M83 and compiled studies of the inner region by Bresolin \& Kennicutt (2002) and Bresolin et al.\ (2005).  The abundance ratio decreases linearly from 12 $+$ log O/H=8.77 in the centre to 12 $+$ log O/H=8.6 at 6 kpc.  

Fig.~\ref{gdr} shows the GDR using a variable X$_\mathrm{CO}$ factor (dashed line).  One notes that the GDR in the inner regions are much reduced and there is now a slight trend towards an increasing GDR with radius.  Some of the values in the inner region are much lower ($\sim$60) than the galactic value of 100-200.  This may not be too surprising given that the metallicity of M83 is super-solar in the inner regions (Bresolin et al.\ 2005), which also implies higher dust masses and subsequently a lower GDR.  This shows the sensitivity of the GDR to an X$_\mathrm{CO}$ factor that varies with metallicity.

The consistency of the GDR over much of the disk beyond the inner regions is in contradiction to several other studies of similarly nearby spiral galaxies.  Recent studies of M99 and M100 (Pohlen et al.\ 2010) and NGC 2403 (Bendo et al.\ 2010b) have shown an increasing GDR with radius. Mu{\~n}oz-Mateo et al.\ (2009) also showed that almost all galaxies in the SINGS sample showed radially increasing GDRs.  However, there were two exceptions.  NGC 3627 showed a decreasing GDR and NGC 2976 showed a fairly consistent GDR.

     %%%%%%%%%%%%%%%%%%%%%%%%%%%%%%%%%%%%%%%%%%%%%%%%%%%%%%%%%%%%%%%%%%%%%%%%%%%%

%%%%%%%%%%%%%%%%%%%%%%%%%%%%%%%%%%%%%%%%%%%%%%%%%%%%%%%%%%%%%%%%%%%%%%%%%%%%

 \section{Conclusions}
 The close proximity of M83 has allowed us to probe in detail the dust temperature and mass distribution using images from PACS and SPIRE.  With its prominent bar and spiral arms we have also been able to compare different environments and see the correspondence of the dust temperature and mass with tracers of the gas and SFR.  We briefly summarize our findings:

\begin{itemize}
\item We find an anti-correlation between $\beta$ and temperature
    with a mean value of $\beta$ of 2.1 $\pm$ 0.1.  However, it is
  possible this anti-correlation is an artifact of noise and line-of-sight temperature variations (Shetty et al.\ 2009a and 2009b).  A comparison of the dust temperature maps made using a constant $\beta$=2 and a variable $\beta$ shows that the most prominent differences are in the inner regions. The dust mass surface density maps are virtually identical and have the same total dust mass within the uncertainties;
\item We find a total dust mass of 4.0 $\pm$ 0.6 $\times$ 10$^{7}$ M$_{\odot}$;
\item  We find that the dust temperature and mass surface density distributions are not spatially coincident.  In particular, while the dust mass surface density correlates extremely well with the molecular gas and SFR, we find that the dust temperature peaks are offset.  In the spiral arms, we find that the dust temperature peaks lie ahead of the dust mass surface density (assuming a trailing spiral structure);
\item The dust temperature is more strongly correlated with the molecular gas than the SFR;
\item  The nuclear and bar-arm transition regions show enhanced dust mass surface densities and temperature.  The bar-arm transition regions lead to the development of a `shoulder' in the radial profile plots.  The position of the shoulder varies between the dust mass surface density and temperature radial profiles;
\item The GDR has a mean value of 84 $\pm$ 4, which is close to the galactic value.  In the inner regions the GDR extends as high as 200; 
\item Using an X$_\mathrm{CO}$ factor that varies with metallicity decreases the GDR in the inner regions.
\end{itemize}

K. Foyle acknowledges helpful conversations with Brent Groves, Frank Israel, Adam Leroy, Giovanni Natale, Karin Sandstrom and Ramin Skibba.  This research was supported by grants from the Canadian Space Agency and the National Science and Engineering Research Council of Canada (PI: C. Wilson).  PACS has been developed by
a consortium of institutes led by MPE (Germany) and including
UVIE (Austria); KU Leuven, CSL, IMEC (Belgium); CEA, LAM
(France); MPIA (Germany); INAF-IFSI/OAA/OAP/OAT, LENS,
SISSA (Italy); IAC (Spain). This development has been supported
by the funding agencies BMVIT (Austria), ESA-PRODEX (Belgium), CEA/CNES (France), DLR (Germany), ASI/INAF (Italy),
and CICYT/MCYT (Spain). SPIRE has been developed by a consortium of institutes led by Cardiff University (UK) and including
Univ. Lethbridge (Canada); NAOC (China); CEA, LAM (France);
IFSI, Univ. Padua (Italy); IAC (Spain); Stockholm Observatory
(Sweden); STFC and UKSA (UK); Imperial College London, RAL, UCL-MSSL, UKATC,
Univ. Sussex (UK); and Caltech, JPL, NHSC, Univ. Colorado
(USA). This development has been supported by national funding agencies: CSA (Canada); NAOC (China); CEA, CNES, CNRS
(France); ASI (Italy); MCINN (Spain); SNSB (Sweden); STFC
(UK); and NASA (USA). HIPE is a joint development by the Herschel Science Ground Segment Consortium, consisting of ESA, the
NASA Herschel Science Center, and the HIFI, PACS and SPIRE
consortia. This research has made use of the NASA/IPAC Extragalactic Database (NED) which is operated by the Jet Propulsion
Laboratory, California Institute of Technology, under contract with
the National Aeronautics and Space Administration.

%\clearpage

%\bibliography{mn-jour}

\appendix
\section{Comparison with Draine \& Li Models}
\begin{figure}
\centering
 \includegraphics[trim=70mm 10mm 30mm 50mm, clip,angle=-90,width=100mm]{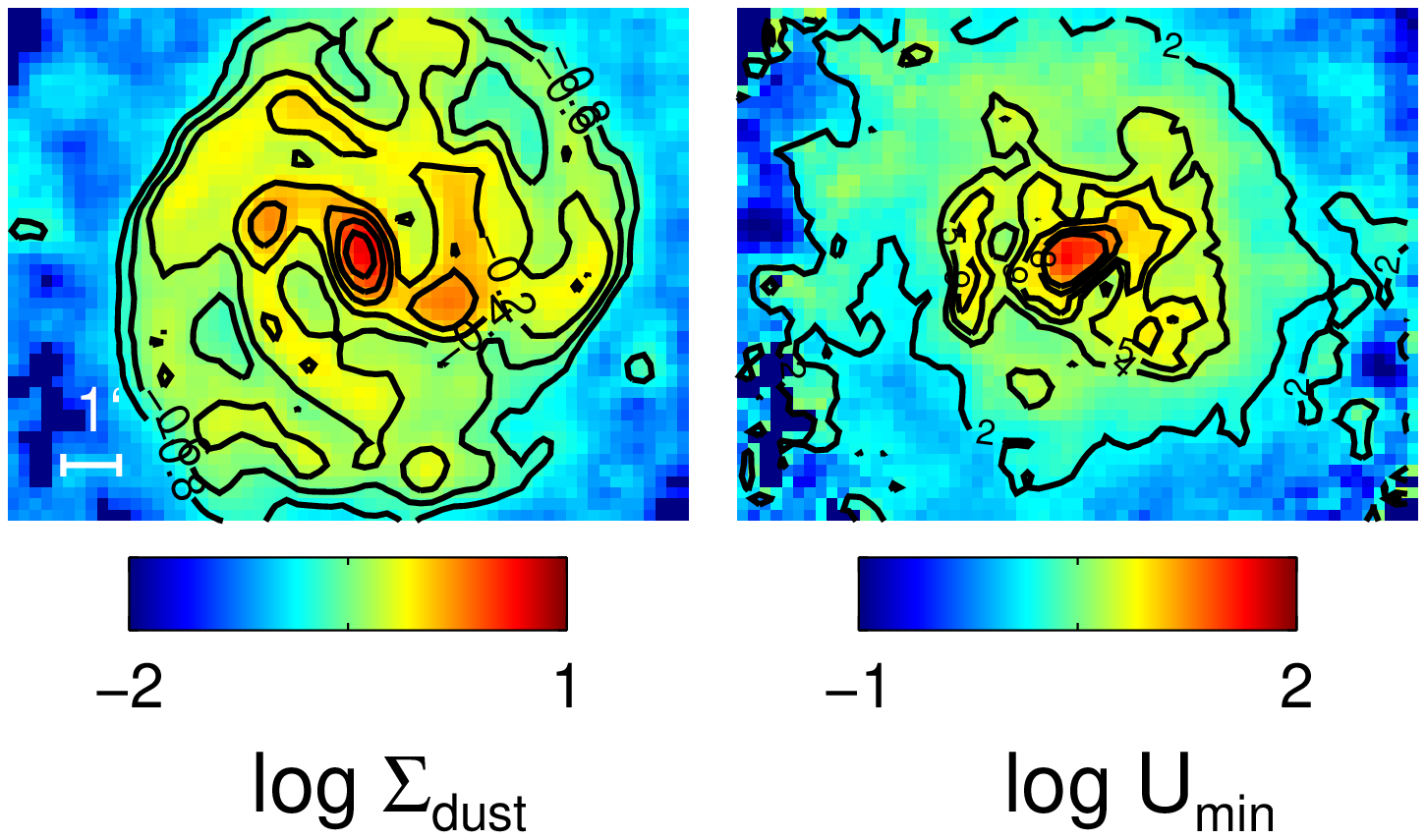}

\caption{Dust mass surface density in log M$_{\odot}$ pc$^{-2}$ (left) and U$_\mathrm{min}$ (right) based on fits using the Draine \& Li (2007) models.  While the dust mass surface density map looks very similar to that made with the modified blackbody fit, the surface densities are all approximately 1.5 times higher (see Fig.~\ref{temp}).  The U$_\mathrm{min}$ map shows similar features to the dust temperature map derived from a modified blackbody fit with a constant $\beta$ (see Fig.~\ref{tempbeta}). }
\label{draine}
\end{figure}

We compare our dust mass estimation using the modified blackbody analysis with that found using the models from Draine \& Li (2007) and the five PACS and SPIRE images. The models are fit using
the two-component emission model put forth by Draine \& Li (2007).  In their model, the mid- through far-infrared emission due to polycyclic aromatic hydrocarbons (PAHs), very small dust grains and carbonaceous dust grains is expressed as a function of the underlying interstellar radiation field (ISRF). The ISRF is represented by two components, one consisting of an emission component modelling a photodissociation
region (PDR) with a range of starlight intensities from U =
U$_\mathrm{min}$ to U=10$^{6}$  and an ISM component with a single radiation field of U=Umin\footnote[1]{We adopt the notation of Draine \& Li (2007).}. The bulk of dust mass is contained in this
latter component. In our least-squares fitting procedure,
U$_\mathrm{min}$ ranged from 0.1 to 25 and we vary the fraction,
$\gamma$, of the PDR to total emission from 10$^{-2}$ to 1. We fix the
PAH fraction, q$_\mathrm{pah}$, to 4.58, common for solar to sub-solar
metallicity spiral galaxies (Draine et al.\ 2007). Errors are computed
by generating 100 realizations of the photometry using a standard
Monte Carlo simulation. We find a total dust mass of 5.3 $\pm$ 1.0
$\times$ 10$^{7}$ M$_{\odot}$, which is factor of $\sim$1.3 greater than a modified blackbody
dust mass with $\beta$=2. The distribution of the dust mass surface density is similar with that found by the modified blackbody fit; in addition the U$_\mathrm{min}$ map is similar to the temperature map using a constant $\beta$.  Fig.~\ref{draine} shows both the dust surface density (left) and U$_\mathrm{min}$ distribution using the Draine \& Li (2007) models.

We find that an enhanced $\gamma$ can fit the photometry
in the nuclear region and the arms, although our variance in these
fitted parameters is larger than the fitted parameters themselves and
thus also gives a result consistent with requiring no PDR region.
U$_\mathrm{min}$ is highest in the nuclear region (mean value in nucleus: 11 $\pm$ 5; maximum value in nucleus: 18 $\pm$ 8) and
enhanced along the spiral arms (mean value in spiral arms: 5 $\pm$1). We also find a similar
offset between the peaks in Umin and the dust mass as we have found between the dust mass and temperature
for the modified blackbody model fits.  

   %%%%%%%%%%%%%%%%%%%%%%%%%%%%%%%%%%%%%%%%%%%%%%%%%%%%%%%%%%%%%%%%%%%%%%%%%%%%

%%%%%%%%%%%%%%%%%%%%%%%%%%%%%%%%%%%%%%%%%%%%%%%%%%%%%%%%%%%%%%%%%%%%%%%%%%%%

\section{Comparison of SFR Tracers}
In this work we used an SFR tracer that was derived from an H$\alpha$ map using a 24~$\mu$m map for compact source corrections (see \S2.3).    This tracer is developed by implicitly assuming that the diffuse 24~$\mu$m emission is not associated with recent star formation and thus should be removed ({\it i.e.} we only use the 24~$\mu$m emission as a compact source correction).  We now examine how these results compare to using an SFR tracer that is a linear combination of two maps and thus includes this diffuse emission.  There are many ways to trace sites of star formation and here we explore two alternative descriptions using a combination of the far-ultraviolet (UV) from GALEX  (Gil de Paz et al.\ 2007) and 24~$\mu$m from Engelbracht et al.\ (2005) as described by Leroy et al.\ (2008) and a linear combination of H$\alpha$ and 24~$\mu$m (Calzetti et al.\ 2007).  We first briefly describe the image rendering for this analysis.

In order to make the SFR$_{UV + 24\mu\text{m}}$, we convolved the UV and 24~$\mu$m maps to the 500~$\mu$m resolution and deprojected them in accordance with Table 2.  The UV and 24~$\mu$m images are combined in the following fashion:

\begin{equation}
\Sigma_{SFR}=(8.1 \times 10^{-2} I_{FUV} +3.2 \times 10^{-3} I_{24}),
\end{equation}
where $\Sigma_{SFR}$ has units of M$_{\sun}$ kpc$^{-2}$ yr$^{-1}$ and the FUV and 24~$\mu$m intensity are each in MJy sr$^{-1}$ (for more details see Appendix A of Leroy et al.\ 2008).

For the SFR$_{H\alpha+ 24\mu\text{m}}$, we convolve each map to the 500~$\mu$m and deproject them in accordance with Table 2.  The H$\alpha$ map is corrected for foreground dust extinction in the Milky Way as previously described in \S2.3.  Rather than applying the 24~$\mu$m correction to H{\small II} regions within the galaxy, we simply combine the two maps using:

\begin{equation}
f_{H\alpha + 24\mu m} = f_{H\alpha} + (0.031)f_{24\mu m}. 
\end{equation}

We then convert this into an SFR using Equation 2 (see \S2.3) and convert this into an SFR surface density.

\begin{figure}
\centering
 \includegraphics[trim=10mm 20mm 10mm 40mm, clip,angle=-90,width=110mm]{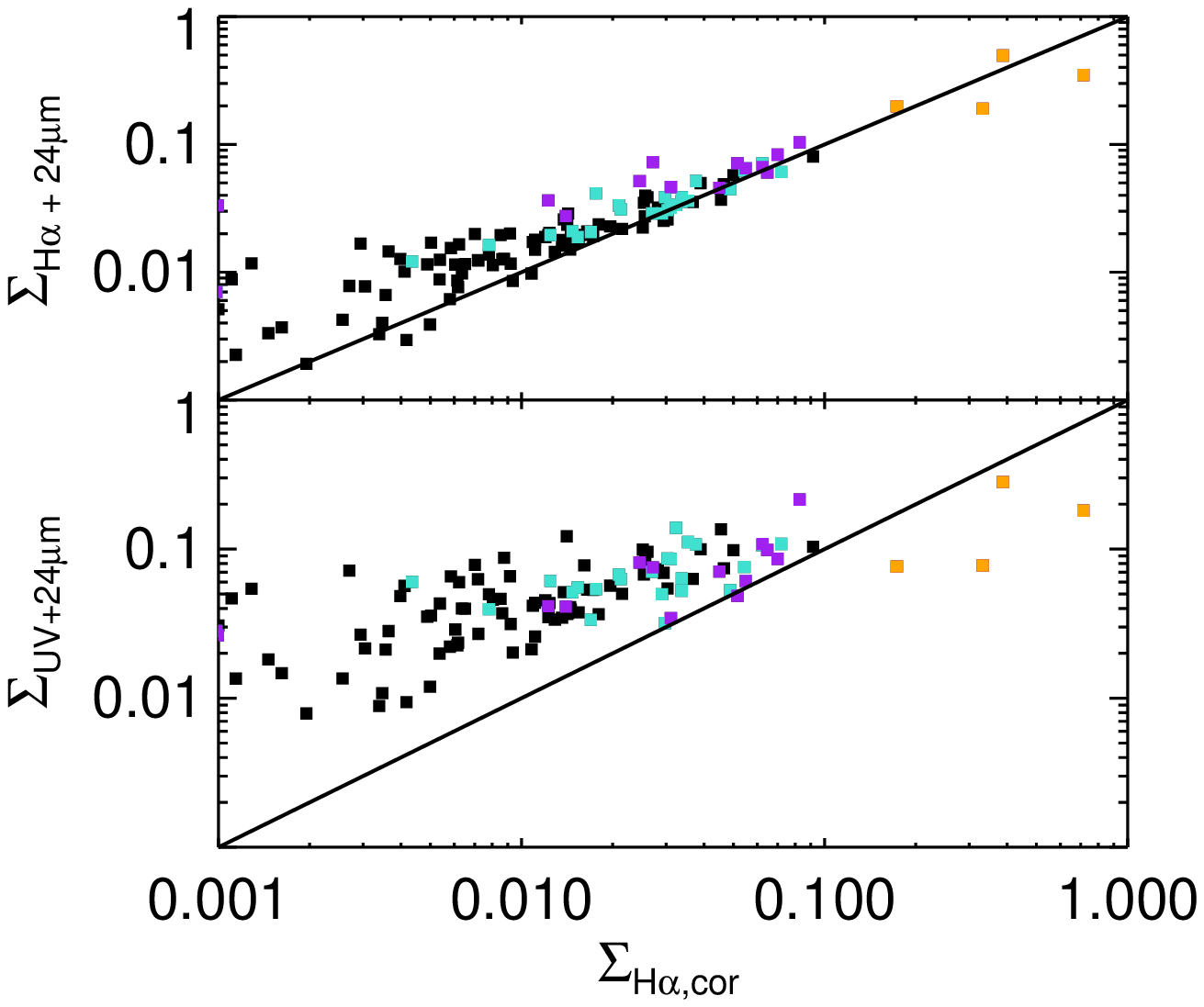}

\caption{Star formation rate as traced by a linear combination of maps (UV + 24~$\mu$m bottom panel and H$\alpha$ + 24~$\mu$m top panel) versus a star formation tracer of H$\alpha_{,cor}$ with a compact source correction using 24~$\mu$m.  Each point represents a pixel of 36 arcsec.}
\label{sfr_comp}
\end{figure}

Fig.~\ref{sfr_comp} provides a direct comparison of the tracers.  On the x-axis we plot the SFR used throughout the text, namely H$\alpha_{,cor}$ with a compact source correction using 24~$\mu$m emission.  For simplicity we refer to this simply as H$\alpha$.  On the y-axis we plot the SFR based on the linear combination of two tracers with the UV + 24~$\mu$m on the bottom panel and H$\alpha$ + 24~$\mu$m on the top panel.  In both cases the linear combination overestimates the SFR at low values of the SFR, with the greatest overestimates being for the UV + 24~$\mu$m.  

As clearly stated in Kennicutt et al.\ (2009), while a linear combination of H$\alpha$ + 24~$\mu$m may be an appropriate description for the SFR as a global quantity, for spatially resolved measures, errors will be introduced.  Due to diffuse dust emission caused by dust heating of older stars, some regions will obtain ``spuriously high SFRs where little or no star formation is taking place" (Kennicutt et al.\ 2009).  For the UV + 24~$\mu$m tracer, there is also the diffuse UV component and the longer timescale for the UV emission ($\sim$100 Myr) in comparison to H$\alpha$ ($\sim$40 Myr) (see Liu et al.\ 2011 for a discussion and comparison of these tracers).
 
\begin{figure*}
\centering
 \includegraphics[trim=10mm 30mm 10mm 40mm, clip,angle=-90,width=150mm]{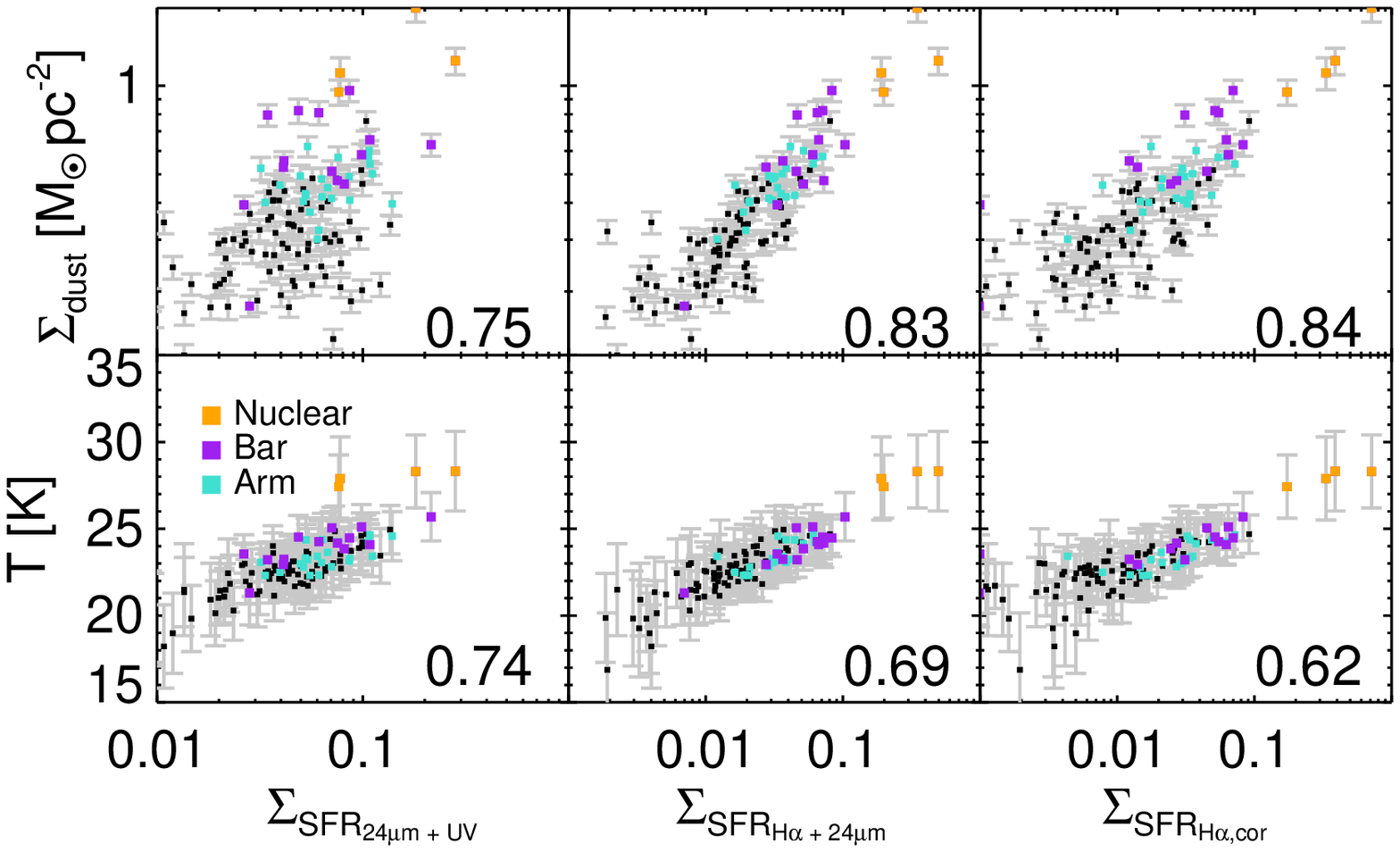}

\caption{Dust mass surface density (upper panel) and dust temperature (lower panel) versus the SFR as traced by a linear combination of UV and 24~$\mu$m (left) and H$\alpha$ and 24~$\mu$m (centre) and H$\alpha_{,cor}$ with a 24~$\mu$m correction for compact sources (right).  Each point represents a pixel of 36 arcsec.  The correlation coefficients are listed in the bottom right of each panel.}
\label{sfr_dust}
\end{figure*}

In Fig.~\ref{sfr_dust} we plot dust mass surface density and dust temperature versus SFR tracer.  We also calculate the correlation coefficients between the different pairs.  In the case of dust surface density we find the correlation coefficients to be 0.75, 0.83 and 0.84 between the UV + 24~$\mu$m, H$\alpha$ + 24~$\mu$m and the H$\alpha_{,cor}$ with a compact source correction respectively.  In other words, the dust surface density is most strongly correlated with the SFR used in this paper.  In the case of the dust temperature we find an opposite ordering of the correlation coefficients with 0.74, 0.69 and 0.62 for UV + 24~$\mu$m, H$\alpha$ + 24~$\mu$m and the H$\alpha_{,cor}$ with a compact source correction, respectively.  This suggests that the tracers composed of linear combinations of maps are capturing a diffuse heating component since they are more strongly correlated with dust temperature rather than dust mass surface density, which is coupled to molecular clouds and subsequent star formation activity.  Thus, the concern that the diffuse 24~$\mu$m emission is not tracing recent star formation seems to be valid based on these findings.  
  %%%%%%%%%%%%%%%%%%%%%%%%%%%%%%%%%%%%%%%%%%%%%%%%%%%%%%%%%%%%%%%%%%%%%%%%%%%%

%%%%%%%%%%%%%%%%%%%%%%%%%%%%%%%%%%%%%%%%%%%%%%%%%%%%%%%%%%%%%%%%%%%%%%%%%%%%
     %%%%%%%%%%%%%%%%%%%%%%%%%%%%%%%%%%%%%%%%%%%%%%%%%%%%%%%%%%%%%%%%%%%%%%%%%%%%

%%%%%%%%%%%%%%%%%%%%%%%%%%%%%%%%%%%%%%%%%%%%%%%%%%%%%%%%%%%%%%%%%%%%%%%%%%%%

\end{document}